\documentclass[12pt]{article}
\usepackage{amssymb}
\usepackage{epsfig}

\catcode`\@=11
\def\numberbysection{\@addtoreset{equation}{section}
        \def\theequation{\thesection.\arabic{equation}}}

\def\beq{\begin{equation}}
\def\eeq{\end{equation}}
\numberbysection
\begin{document}
\begin{titlepage}
\begin{center}
\hfill DFF  2/04/04 \\
\vskip 1.in {\Large \bf  Projectors, matrix models and
noncommutative instantons} \vskip 0.5in P. Valtancoli
\\[.2in]
{\em Dipartimento di Fisica, Polo Scientifico Universit\'a di Firenze \\
and INFN, Sezione di Firenze (Italy)\\
Via G. Sansone 1, 50019 Sesto Fiorentino, Italy}
\end{center}
\vskip .5in
\begin{abstract}
We deconstruct the finite projective modules for the fuzzy
four-sphere, described in a previous paper, and correlate them
with the matrix model approach, making manifest the physical
implications of noncommutative topology. We briefly discuss also
the $U(2)$ case, being a smooth deformation of the celebrated BPST
$SU(2)$ classical instantons on a sphere.
\end{abstract}
\medskip
\end{titlepage}
\pagenumbering{arabic}
\section{Introduction}

The search for instantons on noncommutative manifolds has recently
received growing attention in the physical literature
\cite{18}-\cite{24}. The prospective we take in this paper is to
describe nontrivial configurations in terms of finite projective
modules \cite{8}-\cite{9}, which can be easily constructed in the
noncommutative case \cite{3}-\cite{5}. Then by deconstructing the
projectors we can identify the associated nontrivial connections
satisfying the $Y-M$ equations of motion, following the spirit of
Refs. \cite{6}-\cite{7}. The explicit example we consider is given
in Ref.\cite{4}, where a finite projective module description of
$U(1)$ instantons on a fuzzy four-sphere \cite{10}-\cite{13} has
been presented, based on the Hopf principal fibration $ \pi : S^7
\rightarrow S^4 $. We will briefly treat also the $U(2)$ case
which is physically more interesting since it is a smooth
deformation of the classical BPST $SU(2)$ instanton. Since the
classical limit of the fuzzy four-sphere is more subtle than the
fuzzy two-sphere \cite{14}-\cite{17} , we need to introduce the
problematic related to the fuzzy four-sphere case in the first
sections. The main difference from the fuzzy two-sphere case is
that the coordinates do not form a closed algebra, but we have to
generalize the algebra from five hermitian operators to fifteen
operators. In practice one is obliged to promote as extra
coordinates the commutators of the real coordinates of the sphere.
Therefore when one writes the action of $Y-M$ theory on a fuzzy
four-sphere using a matrix model \cite{10}-\cite{13}, one has
problems in recognizing the classical limit since the derivatives
of the extra coordinates enter into the game and there is no
warranty that their contribution can be decoupled from the
physical coordinates of the sphere $S^4$.

Despite such difficulties we are able to reach some interesting
results. First of all, by introducing a simple link between
projectors and the basic matrix variable $X_i$, we are able to
recognize the class of matrix models for which the projectors
presented in Ref. \cite{4} are solutions to the $Y-M$ equations of
motion, although we have no warranty that the contribution of the
extra coordinates can be decoupled in the classical limit. This is
probably true for the $U(2)$ case, which is a smooth deformation
of the BPST $SU(2)$ instantons, in which case however the link
with matrix models becomes cumbersome.

The last part of the paper is dedicated to recognize the gauge
connection from the projectors \cite{4}, by decomposing the projectors in terms of
more fundamental vector-valued operators. We have to correct a naive decomposition
in terms of oscillators, to avoid a discontinuity problem in the background action,
by dressing this decomposition with quasi-unitary operators. Finally we find a simple
interpretation of the topologically non trivial configurations at a
level of the matrix model \cite{10}-\cite{13}.

\section{Properties of the fuzzy four-sphere}

The fuzzy four-sphere is built in order to match the following two
conditions:

\begin{eqnarray}
& \ & \epsilon^{\mu\nu\lambda\rho\sigma} \hat{x}_\mu \hat{x}_\nu
\hat{x}_\lambda \hat{x}_\rho = C \hat{x}_\sigma \nonumber \\
& \ & \hat{x}_\mu \hat{x}_\mu = R^2 \label{21}
\end{eqnarray}

where $R$ is the radius of the four-sphere. These two conditions
are invariant under a $SO(5)$ group, mixing the coordinates
together.

We solve these two conditions introducing auxiliary finite
matrices $\hat{G}_\mu$ as follows:

\beq \hat{x}_\mu = \rho \hat{G}_\mu \label{22}\eeq

where $\hat{G}_\mu$ can be built from the $n$-fold symmetric
tensor product of the usual Dirac matrices $\gamma_\mu$, see \cite{10}
for details. The finite dimension of $\hat{G}_\mu$, $N$, is
determined to be:

\beq N = \frac{ (n+1)(n+2)(n+3) }{6} \label{23}\eeq

and the constant $C$ must be consistently adjusted.

The matrices $\hat{G}^{(n)}_\mu$ automatically satisfy the
following relations:

\begin{eqnarray}
& \ & \hat{G}^{(n)}_\mu \hat{G}^{(n)}_\mu = n ( n + 4 ) = c
\nonumber \\
& \ & \epsilon^{\mu\nu\lambda\rho\sigma} \hat{G}^{(n)}_\mu
\hat{G}^{(n)}_\nu \hat{G}^{(n)}_\lambda \hat{G}^{(n)}_\rho =
\epsilon^{\mu\nu\lambda\rho\sigma} \hat{G}^{(n)}_{\mu\nu}
\hat{G}^{(n)}_{\lambda\rho} = ( 8 n + 16 ) \hat{G}^{(n)}_\sigma
\label{24}\end{eqnarray}

where

\beq \hat{G}^{(n)}_{\mu\nu} = \frac{1}{2} [ \hat{G}^{(n)}_\mu ,
\hat{G}^{(n)}_\nu ] \label{25}\eeq

Another way to write (\ref{24}) is

\beq \hat{G}^{(n)}_{\mu\nu} = - \frac{1}{2 ( n+2) }
\epsilon^{\mu\nu\lambda\rho\sigma} \hat{G}^{(n)}_{\lambda\rho}
\hat{G}^{(n)}_{\sigma} = - \frac{1}{2(n+2)}
\epsilon^{\mu\nu\lambda\rho\sigma} \hat{G}^{(n)}_\lambda
\hat{G}^{(n)}_\rho \hat{G}^{(n)}_\sigma \label{26} . \eeq

From (\ref{24}) we easily deduce that the conditions (\ref{21}) are met if we pose
$C$ as

\beq C = ( 8 n + 16 ) \rho^3 \ \ \ \ \leftrightarrow \ \ \ \ R^2 =
\rho^2 n ( n+4 ) = \rho^2 c \label{27} . \eeq

We have also the relations

\begin{eqnarray}
& \ & \hat{G}^{(n)}_{\mu\nu} \hat{G}^{(n)}_\nu = 4
\hat{G}^{(n)}_\mu \nonumber \\
& \ & \hat{G}^{(n)}_{\mu\nu} \hat{G}^{(n)}_{\nu\mu} = 4 n ( n + 4
) = 4 c \nonumber \\
& \ & \hat{G}^{(n)}_{\mu\nu} \hat{G}^{(n)}_{\nu\lambda} = c
\delta_{\mu\lambda} + \hat{G}^{(n)}_\mu \hat{G}^{(n)}_\lambda - 2
\hat{G}^{(n)}_\lambda \hat{G}^{(n)}_\mu \label{28} .
\end{eqnarray}

The combination of $\hat{G}^{(n)}_\mu$ and
$\hat{G}^{(n)}_{\mu\nu}$ matrices form a closed $SO(5,1)$ algebra,

\begin{eqnarray}
& \ & [ \hat{G}^{(n)}_\mu , \hat{G}^{(n)}_{\nu\lambda} ] = 2 (
\delta_{\mu\nu} \hat{G}^{(n)}_\lambda - \delta_{\mu\lambda}
\hat{G}^{(n)}_\nu ) \nonumber \\
& \ & [ \hat{G}^{(n)}_{\mu\nu} , \hat{G}^{(n)}_{\lambda\rho} ] = 2
( \delta_{\nu\lambda} \hat{G}^{(n)}_{\mu\rho} + \delta_{\mu\rho}
\hat{G}^{(n)}_{\nu\lambda} - \delta_{\mu\lambda}
\hat{G}^{(n)}_{\nu\rho} - \delta_{\nu\rho}
\hat{G}^{(n)}_{\mu\lambda} ) \label{29} . \end{eqnarray}

Therefore we have to enlarge the coordinates of the sphere from
$5$ to $15$

\begin{eqnarray}
& \ & \hat{x}_\mu = \rho \hat{G}_\mu \nonumber \\
& \ & \hat{w}_{\mu\nu} = i \rho \hat{G}^{(n)}_{\mu\nu} =
\frac{i\rho}{2} [ \hat{G}^{(n)}_\mu , \hat{G}^{(n)}_\nu ]
\label{210} .
\end{eqnarray}

On the fuzzy four-sphere we have the following non-commutativity

\begin{eqnarray}
& \ & [ \hat{x}_\mu, \hat{x}_\nu ] = - 2 i \rho \hat{w}_{\mu\nu}
\nonumber \\
& \ & \epsilon^{\mu\nu\lambda\rho\sigma} \hat{w}_{\mu\nu}
\hat{w}_{\lambda\rho} = - \rho ( 8 n + 16 ) \hat{x}_{\sigma}
\label{211} .
\end{eqnarray}

The classical sphere $S^4$ is obtained as a large $n$ limit
keeping fixed the radius of the sphere $R$, or in other words as a
limit $\rho \rightarrow 0$ with $R$ fixed. From (\ref{211}) we can
see that the coordinates become commuting in this limit:

\begin{eqnarray}
& \ & [ \hat{x}_\mu, \hat{x}_\nu ] = - 2 i \rho \hat{w}_{\mu\nu}
\sim O ( \rho R ) \rightarrow 0 \nonumber \\
& \ & [ \hat{x}_\mu, \hat{w}_{\nu\lambda} ] = 0 \nonumber \\
& \ & [ \hat{w}_{\mu\nu} , \hat{w}_{\lambda\rho} ] = 0 \label{212}
.
\end{eqnarray}

The non-commutativity is caused by the presence of the extra
coordinates $\hat{w}_{\mu\nu}$; in practice we can state that an
extra fuzzy two-sphere is attached to every point of fuzzy
four-sphere. To realize that let's diagonalize the matrix
$\hat{x}_5 = \rho \hat{G}_5$. Then there is a subalgebra $SU(2)
\times SU(2)$  generated by $\hat{G}_{\mu\nu} ( \mu, \nu = 1,...,
4 )$ of the $SO(5)$ algebra generated by $\hat{G}_{\mu\nu} ( \mu,
\nu = 1, ... , 5  )$, commuting with $\hat{x}_5$:

\begin{eqnarray}
& \ & [ \hat{N}_i , \hat{N}_j ] = i \epsilon_{ijk} \hat{N}_k
\nonumber \\
& \ & [ \hat{M}_i , \hat{M}_j ] = i \epsilon_{ijk} \hat{M}_k
\nonumber \\
& \ & [ \hat{N}_i , \hat{M}_j ] = 0 \label{213}
\end{eqnarray}

where

\begin{eqnarray}
& \ & \hat{N}_1 = - \frac{i}{4} ( \hat{G}_{23} - \hat{G}_{14} ) \
\ \ \ \ \ \ \ \ \ \hat{M}_1 = - \frac{i}{4} ( \hat{G}_{23} +
\hat{G}_{14} ) \nonumber \\
& \ & \hat{N}_2 = \frac{i}{4} ( \hat{G}_{13} + \hat{G}_{24} ) \ \
\ \ \ \ \ \ \ \ \hat{M}_2 = \frac{i}{4} ( \hat{G}_{13} -
\hat{G}_{24} ) \nonumber \\
& \ & \hat{N}_3 = - \frac{i}{4} ( \hat{G}_{12} - \hat{G}_{34} ) \
\ \ \ \ \ \ \ \ \ \hat{M}_3 = - \frac{i}{4} ( \hat{G}_{12} +
\hat{G}_{34} ) \label{214} .
\end{eqnarray}

Conversely, $\hat{G}_{\mu\nu} ( \mu, \nu = 1, ..., 4)$ can be
written as:

\begin{eqnarray}
& \ & \hat{G}_{23} = 2i ( \hat{N}_1 + \hat{M}_1 ) \ \ \ \ \ \ \ \
\ \ \  \hat{G}_{14} = - 2i ( \hat{N}_1 - \hat{M}_1 ) \nonumber \\
& \ & \hat{G}_{13} = - 2i ( \hat{N}_2 + \hat{M}_2 ) \ \ \ \ \ \ \
\ \ \ \  \hat{G}_{24} = - 2i ( \hat{N}_2 - \hat{M}_2 ) \nonumber \\
& \ & \hat{G}_{12} = 2i ( \hat{N}_3 + \hat{M}_3 ) \ \ \ \ \ \ \ \
\ \ \  \hat{G}_{34} = - 2i ( \hat{N}_3 - \hat{M}_3 ) \label{215} .
\end{eqnarray}

The Casimir of every $SU(2)$ algebra is computed as follows:

\begin{eqnarray}
& \ & \hat{N}_i \hat{N}_i = \frac{1}{16} ( n + G_5 ) ( n + 4 + G_5
) \nonumber \\
& \ & \hat{M}_i \hat{M}_i = \frac{1}{16} ( n - G_5 ) ( n + 4 - G_5
) \label{216} . \end{eqnarray}

Fixing the value of $\hat{G}_5 = G_5$ the $SU(2)$ algebra generated by $\hat{N}_i$ is
realized by a $( \frac{ n + G_5 + 2 }{2} )$ dimensional representation, while the
$SU(2)$ algebra generated by $\hat{M}_i$ is realized by a
$( \frac{ n - G_5 + 2 }{2} )$ dimensional representation, resulting in
$\frac{ ( n + G_5 + 2 )( n - G_5 + 2 )}{4}$ possible eigenvalues. If we sum
up the contributions of $ G_5 = - n, -n+2 , ...., n-2, n $ we end
up with the dimension $N$ of the matrix.

We notice that an $SU(2)$ algebra decouples at the north pole, $G_5 = n$, since
the Casimir of $\hat{N}_i$ and $\hat{M}_i$ are given by:

\begin{eqnarray}
& \ & \hat{N}_i \hat{N}_i = \frac{ n ( n + 2 )}{4} \nonumber \\
& \ & \hat{M}_i \hat{M}_i = 0 \label{217} .
\end{eqnarray}

We have reached the result that a fuzzy two-sphere, given by the
$(n+1)$-dimensional representation of $SU(2)$ is attached to the
north pole. The radius of the fuzzy two-sphere $\sigma^2 = \rho^2
\frac{n(n+4)}{4}$ is comparable with the radius of the fuzzy
four-sphere, $ R^2 = \rho^2 n ( n + 4 )$.

By using the $SO(5)$ symmetry we can generalize this observation
from the north pole to every point of the fuzzy four-sphere. This
fuzzy two-sphere is a sort of internal space, in the sense that
the fields on the fuzzy four-sphere carry internal quantum numbers
corresponding to the $SU(2)$ angular momentum and therefore these
extra degrees of freedom have the nice interpretation as spin.

\section{Matrix model for the fuzzy four-sphere}

To introduce a gauge theory on a fuzzy four-sphere we consider the
following matrix model:

\beq S = -  \frac{1}{g^2} Tr \left[ \frac{1}{4} [ X_\mu, X_\nu ] [
X_\mu, X_\nu ] + \frac{k}{5} \epsilon^{\mu\nu\lambda\rho\sigma}
X_\mu X_\nu X_\lambda X_\rho X_\sigma \right] \label{31}\eeq

where the indices $\mu, \nu, ...,\sigma$ take the values $1,...,5$
and are contracted with the Euclidean metric.
$\epsilon^{\mu\nu\lambda\rho\sigma}$ is the $SO(5)$ invariant
totally antisymmetric tensor. $X_\mu$ are hermitean $N \times N$
matrices and $k$ is a dimensional constant depending on $N$. The
second term is known as Myers term and it has a consistent brane
interpretation \cite{13}.

This model has the global $SO(5)$ symmetry and the following
unitary symmetry

\beq X_\mu = U^{\dagger} X_\mu U \ \ \ \ \ \ \ U U^{\dagger} =
U^{\dagger} U = 1 \label{32} \eeq

apart from an extra translational symmetry $X_\mu \rightarrow
X_\mu + c_\mu 1$.

The constant $k$ is determined by the conditions that the matrix
model (\ref{31}) has as a classical solution the fuzzy four-sphere:

\beq X_\mu = \hat{x}_\mu = \rho \hat{G}^{(n)}_\mu \label{33} .
\eeq

Since the equations of motion of the action (\ref{31}) are of the form

\beq [ X_\nu, [ X_\mu, X_\nu ]] + k
\epsilon^{\mu\nu\lambda\rho\sigma} X_\nu X_\lambda X_\rho X_\sigma
= 0  \label{34} \eeq
the constant $ k $ is determined in terms of
the label $n$:

\beq k = \frac{2}{\rho ( n+2 )} \label{35} . \eeq

It is also possible to introduce a Y-M action with mass term
having the fuzzy four-sphere as a classical solution:

\beq S = - \frac{1}{g^2} Tr \left[ \frac{1}{4} [ X_\mu, X_\nu ][
X_\mu, X_\nu ] + 8 \rho^2 X_\mu X_\mu  \right] \label{36} . \eeq

The construction of the noncommutative gauge theory on the fuzzy
four-sphere is obtained by expanding the general matrices $X_\mu$
around the classical background $\hat{x}_\mu$ :

\beq X_\mu = \hat{x}_\mu + \rho R \hat{a}_\mu \label{37} . \eeq

The functional space on which the fields $X_\mu$ live is
determined by the background. While the fields on the sphere can
be developed in terms of the spherical harmonics, the field theory on a fuzzy sphere
is realized by truncating the angular momentum with a cutoff parameter.
Many papers deal with such construction
for higher dimensional fuzzy spheres, see for example \cite{25}. In four dimensions the
basis is classified by irreducible representations of $SO(5)$. The
corresponding $SO(5)$ Young diagram is labelled by two labels
$(r_1,r_2)$. Only the representations with $r_2 = 0$ correspond to
the classical sphere ( $w_{\mu\nu} = 0$ ).

The fuzzy four-sphere functional space is therefore obtained by
the $SO(5)$ irreducible representation with the cutoff $r_1 \leq
n$, i.e. with $ 0 \leq r_2 \leq r_1 \leq n $. By summing the
dimensions of all these irreducible representations we obtain the
square of $N$, the rank of the matrix $\hat{G}_\mu$.

Therefore a general $N \times N$ matrix $\hat{a}_\mu$ can be
developed in an abstract form:

\beq \hat{a} ( \hat{x}, \hat{w} ) = \sum^{n}_{r_1 = 0}
\sum^{r_1}_{r_2 = 0} \sum_{m_i} \ a_{r_1 r_2 m_i} \hat{Y}_{r_1 r_2
m_i } ( \hat{x}, \hat{w} ) \label{38}\eeq

where $m_i$ denote all the relevant quantum numbers.
In the case of $\hat{w}_{\mu\nu} = 0$, $\hat{Y}_{r_1 m_i}$ are the
usual spherical harmonics. However in the fuzzy four-sphere case
we need to assume that the fields depend also on the extra
coordinates $\hat{w}_{\mu\nu}$.

The symbol corresponding to the matrix $\hat{a} ( \hat{x}, \hat{w}
)$ is easily obtained as

\beq a( x, w) = \sum^{n}_{r_1 = 0} \sum^{r_1}_{r_2 = 0} \sum_{m_i}
\ a_{r_1 r_2 m_i} Y_{r_1 r_2 m_i } ( x , w ) \label{39}\eeq

and the product of matrices can be mapped to a noncommutative and
associative star product of symbols. The non-commutativity of the
star product is produced by the existence of $\hat{w}_{\mu\nu}$.

To define the action of a noncommutative gauge theory on a fuzzy
four-sphere we need to introduce derivative operators such as

\begin{eqnarray}
& \ & Ad ( \hat{G}_\mu ) \ \ \ \rightarrow - 2 i \left( w_{\mu\nu}
\frac{ \partial}{\partial x_\nu} - x_\nu \frac{
\partial}{\partial w_{\mu\nu}} \right) \nonumber \\
& \ & Ad ( \hat{G}_{\mu\nu} ) \ \ \ \rightarrow 2 \left( x_\mu
\frac{ \partial}{\partial x_\nu} - x_\nu \frac{
\partial}{\partial x_\mu } - w_{\mu\lambda} \frac{ \partial}{\partial w_{\lambda\nu}}
+  w_{\nu\lambda} \frac{ \partial}{\partial w_{\lambda\mu}}
\right) \label{310} .
\end{eqnarray}

In the case of $Ad ( \hat{G}_{\mu\nu} )$ we can isolate the first two terms
corresponding to the orbital parts while the last two have the meaning of
isospin parts.

To make clear that the fields $\hat{a} ( \hat{x}, \hat{w} )$ are
spin-dependent representations of the $SO(4)$ internal Lorentz
group generated by $G_{ab}, (a,b = 1,...,4)$ , we define its
action as :

\begin{eqnarray}
& \ & e^{ i \hat{G}_{ab} \omega_{ab} } \hat{a}  ( \hat{x}, \hat{w}
) e^{ - i \hat{G}_{ab} \omega_{ab} } \simeq \hat{a}  ( \hat{x},
\hat{w} ) + i \omega_{ab} Ad ( G_{ab} ) \hat{a}  ( \hat{x},
\hat{w} ) \nonumber \\
& \ & \rightarrow a( x, w) + 2 i \omega_{ab} \left( x_a
\frac{\partial}{\partial x_b} - x_b \frac{\partial}{\partial x_a}
- w_{ac} \frac{\partial}{\partial w_{cb}} + w_{bc}
\frac{\partial}{\partial w_{ca}} \right) a(x,w) \label{311} .
\end{eqnarray}

The last term shows that the fields $a(x,w)$ have spin angular
momentum, taking only integer values and with a maximum spin
limited by the dimension of the matrices $\hat{N}_i$, which is
$(n+1)$.

To isolate the spin $m$ ( $m < n$ ) contribution, it is necessary
to develop the field $a (x,w)$ ,for example at the north pole, in
terms of the coordinates $N_i$:

\beq a(x,w) = a (x,0) + N_{i_1} \frac{ \partial a(x,N) }{ \partial
N_{i_1} } |_{N=0} + .... + \frac{1}{n!} N_{i_1} N_{i_2} ...
N_{i_n} \frac{
\partial^n a(x,N) }{
\partial N_{i_1} \partial N_{i_2}.... \partial N_{i_n} } |_{N=0}
\label{312} . \eeq

Each term in this development has a definite spin, i.e. $(m+1)$-th
term represents an $m$ spin field. This $N$ dependence on the
internal spin produces non-commutativity. It is also possible to remove
the fuzzy two-sphere ( $N_i$ ) from the fuzzy four-sphere, however the product of
fields, although commutative, becomes non-associative.

The gauge symmetry of the non-commutative gauge theory is a direct consequence
of the unitary symmetry of the matrix model. By taking an infinitesimal
transformation $U \simeq 1 + i \hat{\lambda} $, a fluctuation
around the fixed background transforms, similarly to a gauge field, as

\beq \delta \hat{a}_\mu ( \hat{x}, \hat{w} ) = - \frac{i}{R} [
\hat{G}_\mu , \hat{\lambda} ( \hat{x}, \hat{w} ) ] + i [
\hat{\lambda} ( \hat{x}, \hat{w} ), \hat{a}_\mu ( \hat{x},\hat{w}
) ] \label{313} \eeq

and for the corresponding symbol

\beq \delta a_\mu ( x, w ) = \frac{2}{R} \left( w_{\mu\nu}
\frac{\partial}{\partial x_\nu} - x_\nu \frac{\partial}{\partial
w_{\mu\nu}} \right) \lambda ( x, w ) + i [ \lambda ( x, w ), a_\mu
( x , w ) ]_{*} \label{314} . \eeq

By developing $\hat{\lambda}$ as

\beq \hat{\lambda} = \lambda_0 + \lambda^\mu \hat{G}_\mu +
\lambda^{\mu\nu} \hat{G}_{\mu\nu} + 0 ( G^2 ) \label{315}\eeq

this gauge transformation contains many extra degrees of freedom which have no
equivalence in a standard gauge theory.

The integration in the classical gauge theory is replaced by the
trace in the action of the matrix model. However in the
correspondence one has to take into account the presence of the
extra internal two-dimensional space $N_i$, apart from the fuzzy
four-sphere.

Finally the Laplacian on the sphere has two possible extensions on
the fuzzy four-sphere, $Ad ( \hat{G}_{\mu\nu})^2$ the quadratic
Casimir of $SO(5)$ and $Ad ( \hat{G}_{\mu})^2$. The natural choice
for a matrix model, in which we develop the matrices $X_\mu$
around the background $ \hat{x}_\mu = \rho \hat{G}_\mu $, is given
by $Ad ( \hat{G}_{\mu})^2$, whose action is given by

\begin{eqnarray}
& \ & \frac{1}{4} \left[ \frac{\hat{G}_\mu}{R} , \left[
\frac{\hat{G}_\mu}{R} , . \right] \right] =
\frac{\partial^2}{\partial x_\mu \partial x_\mu} - \frac{x_\mu
x_\nu}{R^2} \frac{\partial^2}{\partial x_\mu \partial x_\nu}
 - \frac{4 x_\mu}{R^2} \frac{\partial}{\partial x_\mu }
 \nonumber \\
 & \ & - \frac{2 w_{\mu\nu} x_\lambda}{R^2}
 \frac{\partial^2}{\partial x_\nu \partial w_{\mu\lambda}}
 + \frac{x_\nu x_\lambda}{R^2} \frac{\partial^2}{\partial
 w_{\mu\nu} \partial w_{\mu\lambda}} -
 \frac{w_{\mu\lambda}}{R^2} \frac{\partial}{\partial
 w_{\mu\lambda}} \label{316} .
 \end{eqnarray}

We recognize the usual Laplacian on a classical four-sphere in the first three terms.
The action of the two Laplacians on a spherical harmonics is as follows:

\begin{eqnarray}
& \ & \frac{1}{4} \left[ \hat{G}_\mu , \left[ \hat{G}_\mu ,
\hat{Y}_{r_1 r_2} \right] \right] = ( r_1 ( r_1 + 3 ) - r_2 ( r_2
+ 1 ) ) \hat{Y}_{r_1, r_2} \nonumber
\\
& \ & - \frac{1}{8} \left[ \hat{G}_{\mu\nu} , \left[
\hat{G}_{\mu\nu} , \hat{Y}_{r_1 r_2} \right] \right] = ( r_1 ( r_1
+ 3 ) + r_2 ( r_2 + 1 ) ) \hat{Y}_{r_1, r_2} \label{317} .
\end{eqnarray}

\section{Projectors for the fuzzy four-sphere}

A general procedure to characterize instantons configurations on
the classical sphere $S^4$ is starting from the algebra of $N
\times N$ matrices whose entries are elements of the smooth
function algebra $C^{\infty} (S^4)$ on the base space $S^4$, i.e.
$M_N ( C^{\infty} ( S^4 ))$. The section module of the bundle on
which the instanton lives can be identified with the action of a
global projector $ p \in M_N ( C^{\infty} ( S^4 ))$ on the trivial
module ${( C^{\infty} ( S^4))}^N$, i.e. the right module $ p {(
C^{\infty} ( S^4))}^N $, where an element of $ {( C^{\infty} (
S^4))}^N $ is simply the vector

\beq || f >> = \left( \begin{array}{c} f_1 \\ ... \\ f_N
\end{array}\right) \label{41}\eeq

with $ f_1, ..., f_N $ elements of $ C^{\infty} ( S^4 ) $.

In Ref. \cite{4} we found $U(1)$ instantons configurations for the fuzzy
four-sphere. These have been obtained generalizing the
construction of the non-commutative monopoles, based on the Hopf
fibration $\pi : S^3 \rightarrow S^2$ to the case $\pi : S^7
\rightarrow S^4$.

Briefly speaking, the Hopf fibration $\pi : S^7 \rightarrow S^4$
is a classical map from four complex coordinates $a_i$, constrained to live in $S^7$,
to five real coordinates $x_i$, constrained to live in $S^4$:

\begin{eqnarray}
& \ & x_1 = \rho ( \alpha_1 + \overline{\alpha}_1 ) \ \ \ \ \ \ \
\ x_2 = i \rho ( \alpha_1 - \overline{\alpha}_1 ) \nonumber \\
& \ & x_3 = \rho ( \alpha_2 + \overline{\alpha}_2 ) \ \ \ \ \ \ \
\ x_4 = i \rho ( \alpha_2 - \overline{\alpha}_2 ) \nonumber \\
& \ & x_5 = \rho ( a_0 \overline{a}_0 + a_1 \overline{a}_1 - a_2
\overline{a}_2 - a_3 \overline{a}_3 ) \nonumber \\
& \ & \alpha_1 = a_0 \overline{a}_2 + a_3 \overline{a}_1 \nonumber
\\
& \ & \alpha_2 = a_0 \overline{a}_3 - a_2 \overline{a}_1 \nonumber
\\
& \ & \sum_i x^2_i = \rho^2 \ \ \ \ \ \ \ \ \sum_i | a_i |^2 = 1
\label{42} .
\end{eqnarray}

The idea proposed in Ref. \cite{4} is that promoting the complex coordinates $a_i$
to four oscillators

\beq [ a_i, a^{\dagger}_j ] = \delta_{ij} \ \ \ \ \ \ \ \ \ [ a_i,
a_j ] = 0 \label{43}\eeq

the coordinates $\hat{x}_i$ are part of an algebra, coinciding with the fuzzy four-sphere
algebra. For example the $SU(2) \times SU(2)$ subalgebra made by
$N_i$ and $M_i$ can be represented in terms $a_i$ as follows:

\begin{eqnarray}
& \ & \hat{N}_3 = \frac{1}{2} ( a_3 a^{\dagger}_3 - a_2
a^{\dagger}_2 ) \ \ \ \ \ \ \ \ \ \hat{M}_3 = \frac{1}{2} ( a_0
a^{\dagger}_0 - a_1 a^{\dagger}_1 ) \nonumber \\
& \ & \hat{N}_{+} = \hat{N}_1 + i \hat{N}_2 = a_2 a^{\dagger}_3 \
\ \ \ \ \ \ \ \ \ \ \ \hat{M}_{+} = \hat{M}_1 + i \hat{M}_2 = a_1
a^{\dagger}_0 \nonumber \\
& \ & \hat{N}_{-} = \hat{N}_1 - i \hat{N}_2 = a_3 a^{\dagger}_2 \
\ \ \ \ \ \ \ \ \ \ \ \hat{M}_{-} = \hat{M}_1 - i \hat{M}_2 = a_0
a^{\dagger}_1 \label{44} .
\end{eqnarray}

It is possible to define a total number operator $\hat{N}$, whose
eigenvalue corresponds exactly with the label $n$ of the
representation $\hat{G}_{\mu}^{(n)}$, introduced in formula
(\ref{23}):

\beq \hat{N} = a^{\dagger}_0 a_0 + a^{\dagger}_1 a_1 +
a^{\dagger}_2 a_2 + a^{\dagger}_3 a_3 \ \ \ \ \ \ \ \ \ \ \
\hat{N} \rightarrow n \label{45} . \eeq

In terms of $\hat{N}$ the Casimir for $\hat{x}^2_i$ is:

\beq \sum_i \hat{x}^2_i = \rho^2 \hat{N} ( \hat{N} + 4 ) = R^2
\label{46} . \eeq

The idea of Ref. \cite{4} is to construct the projective modules for the $k$-instanton,
starting from a vector whose entries belong to the oscillator algebra (\ref{43})

\beq | \psi_k > = N_k ( \hat{N} ) \left( \begin{array}{c} {(a_0)}^k \\
....... \\ \sqrt{ \frac{ k! }{ i_1 ! i_2 ! i_3 ! ( k - i_1 - i_2 -
i_3 )! } } {(a_0)}^{k-i_1-i_2-i_3} {(a_1)}^{i_1} {(a_2)}^{i_2} {(a_3)}^{i_3} \\
....... \\ {(a_1)}^k \end{array} \right) \label{47} . \eeq

where $0 \leq i_1 \leq k$, $0 \leq i_2 \leq k-i_1$, $0 \leq i_3
\leq k-i_1-i_2$.

In total the number of entries of the vector $| \psi_k >$ is given by summing on
$i_1, i_2, i_3$ as follows:

\beq N_k = \frac{ ( k+1 ) ( k+2 ) ( k+3 ) }{6} \label{48}\eeq

which can be identified with the matrix $\hat{\Gamma}^{(k)}_\mu$ ( i.e. posing $n=k$ ).

Fortunately the normalization condition for $ | \psi_k > $ can be solved by a function
function $\hat{N}_k$ of the number operator $\hat{N}$ as follows:

\begin{eqnarray}
& \ & < \psi_k | \psi_k > = 1 \nonumber \\
& \ & N_k = N_k ( \hat{N} ) = \frac{1}{\sqrt{ \prod^{k-1}_{i=0} (
\hat{N} - i + k ) }} = \frac{1}{\sqrt{ \prod^{k-1}_{i=0} ( n - i +
k ) }} \label{49} . \end{eqnarray}

The projector for the $k$-instanton on the fuzzy four-sphere is
defined as

\beq P_k = | \psi_k >< \psi_k | \ \ \ \ \ \ \ \ \ P^2_k = P_k \ \
\ P^{\dagger}_k = P_k \label{411} . \eeq

In the ket-bra product there appear only entries commuting with the number operator
$\hat{N}$, and therefore those are polynomial functions of the sixteen combinations
$a_i a^{\dagger}_j$, which produce the fuzzy four-sphere algebra and the number
operator, equal to its eigenvalue $n$. Therefore we fulfill the requirement that the
projector $P_k$ has as entries the elements of the basic operator algebra of the theory.

We can verify that the trace of the projector $P_k$ is always a positive integer, as it
should be

\begin{eqnarray}
& \ & Tr \  P_k = \frac{ ( n+k+1 ) ( n+k+2 ) ( n+k+3 ) }{ ( n+1 )
(
n+2 ) ( n+3 ) } \ Tr 1 \nonumber \\
& \ & = \frac{ ( n+k+1 ) ( n+k+2 ) ( n+k+3 ) }{ 6 } < \frac{ ( n+1
) ( n+2 ) ( n+3 ) }{ 6 } \frac{ ( k+1 ) ( k+2 ) ( k+3 ) }{ 6 } =
Tr 1_P \nonumber \\
& \ & \label{412} \end{eqnarray}

where $1_P$ is the identity projector.

With the knowledge of the vector-valued operator $ < \psi_k | $ it is possible
to compute in principle the corresponding $1$-form connection for a $k$-instanton

\beq A^\nabla_k = < \psi_k | d | \psi_k > \label{410} . \eeq

To build the $(-k)$-instanton it is enough to take the analogous
of the vector $| \psi_k >$ with adjoint entries, apart from a new
normalization function $ N_k ( \hat{N} ) $. In fact we
consider the vector

\beq | \psi_{-k} > = N_k ( \hat{N} ) \left( \begin{array}{c}
{(a^{\dagger}_0)}^k \\
....... \\
\sqrt{ \frac{ k! }{ i_1 ! i_2 ! i_3 ! ( k - i_1 - i_2 -
i_3 )! } } {(a^{\dagger}_0)}^{k-i_1-i_2-i_3}
{(a^{\dagger}_1)}^{i_1} {(a^{\dagger}_2)}^{i_2}
{(a^{\dagger}_3)}^{i_3} \\
....... \\
{(a^{\dagger}_1)}^k \end{array} \right) \label{413} . \eeq

Again we obtain the good property that the normalization condition is solved by a function
$N_k$ of the number operator:

\begin{eqnarray}
& \ & < \psi_{-k} | \psi_{-k} > = 1 \nonumber \\
& \ & N_k = N_k ( \hat{N} ) = \frac{1}{\sqrt{ \prod^{k-1}_{i=0} (
\hat{N} + i + 4 - k ) }} = \frac{1}{\sqrt{ \prod^{k-1}_{i=0} ( n +
i + 4 - k ) }} \label{414} .
\end{eqnarray}

The corresponding projector for the $(-k)$-instanton is

\beq P_{-k} = | \psi_{-k} >< \psi_{-k} | \ \ \ \ \ \ \ \ \  k < n+
4 \label{415} . \eeq

We have to be careful with the trace of this projector

\begin{eqnarray}
& \ & Tr \ P_{-k} = \frac{ ( n-k+1 ) ( n-k+2 ) ( n-k+3 ) }{ ( n+1
)
( n+2 ) ( n+3 ) } \ Tr \ 1 = \nonumber \\
& \ & = \frac{ ( n-k+1 ) ( n-k+2 ) ( n-k+3 ) }{ 6 } < \ Tr \ 1_P
\label{416} \end{eqnarray}

since it is definite positive if and only if

\beq k < n + 1 \label{417} . \eeq

For the special cases $ k = n+1, n+2, n+3 $, $P_{-k}$ is simply
the null projectors.

\section{Projectors for the $SU(2)$ instantons}

A nice paper \cite{8} deals with a finite projective module description
of the non-trivial $SU(2)$ gauge configurations on the sphere
$S^4$. In the case of $SU(2)$ instanton we must take care of the
vector space underlying the theory, i.e. the quaternion field $H$.
Let's define $ A_H = C^{\infty} ( S^4, H ) $ the algebra of the
smooth functions taking values in $ H $ on the base space $S^4$.
The projector $ p \in M_N ( A_H ) $ for the quaternion valued
functions can be built using the principal Hopf fibration $ \pi :
S^7 \rightarrow S^4 $ on the sphere. It is convenient to introduce
also $ B_H = C^{\infty} ( S^7, H ) $ the algebra of smooth
functions with values in $H$ on the total base space $S^7$.

The projector can be written as

\beq p = | \psi > < \psi | \label{51}\eeq

with

\beq | \psi > = \left( \begin{array}{c} \psi_1 \\ ... \\ \psi_N
\end{array}\right) \label{52}\eeq

a vector valued function on $S^7$, i.e. an element of $ {( B_H
)}^N $. Imposing that the vector valued function $ | \psi > $ is
normalized, i.e.

\beq < \psi | \psi > = 1 \label{53}\eeq

we obtain that $p$ is a projector since

\beq p^2 = | \psi > < \psi | \psi > < \psi | = p \ \ \ \ \
p^{\dagger} = p \label{54} . \eeq

We have already introduced the principal Hopf fibration $ SU(2)
\simeq Sp(1) \  \pi : S^7 \rightarrow S^4 $ on the
four-dimensional sphere, but to prepare the $SU(2)$ case we need
to realize it in terms of a couple of quaternions, instead of four
oscillators:

\beq S^7 = \{ ( a, b ) \in H^2 , | a |^2 + | b |^2 = 1 \}
\label{55}\eeq

with right action

\begin{eqnarray}
& \ & S^7 \times Sp(1) \rightarrow S^7 \ \ \ \ \ \ \  ( a, b ) w =
( aw, bw ) \nonumber \\
& \ & w \in Sp(1) \ \ \ \ \leftrightarrow \ \ \ \ w \overline{w} =
1  \label{56} . \end{eqnarray}

The right action (\ref{56}) respects the $S^7$ constraint. In terms of the quaternions
$ a, b$ the fiber bundle projector $ \pi : S^7 \rightarrow S^4 $, the Hopf
fibration, is realized as $ \pi (a,b) = ( x_1, x_2,
x_3, x_4, x_5 ) $

\begin{eqnarray}
& \ & x_1 = a \overline{b} + b \overline{a} \nonumber \\
& \ & \xi = a \overline{b} - b \overline{a} = - \overline{\xi} \nonumber \\
& \ & x_5 = | a |^2 - | b |^2 \nonumber \\
& \ & \sum_{\mu = 1}^5  {( x_\mu )}^2 = {( | a |^2 + | b |^2 )}^2
= 1 \label{57} .
\end{eqnarray}

The basic $Sp(1)$ invariant functions on $S^7$ are obtained by inverting
(\ref{57})

\begin{eqnarray}
& \ & | a |^2 = \frac{1}{2} ( 1 + x_5 ) \nonumber \\
& \ & | b |^2 = \frac{1}{2} ( 1 - x_5 ) \nonumber \\
& \ & a \overline{b} = \frac{1}{2} ( x_1 + \xi ) \label{58}
\end{eqnarray}

from which we can build a generic ( polynomial ) invariant function on $S^7$
as a function of these variables.

The projector for $k=1$ instanton is obtained by considering the following
ket valued function:

\beq | \psi > = \left( \begin{array}{c} a \\ b \end{array} \right)
\label{59} \eeq

satisfying the normalization condition $ < \psi | \psi > = | a |^2 + | b |^2 = 1 $ on $S^7$.

We can define a projector in $ M_2 ( A_H )$ as

\beq p = | \psi > < \psi | = \left( \begin{array}{cc} | a |^2 & a
\overline{b} \\ b \overline{a} & | b |^2
\end{array} \right) = \frac{1}{2} \left( \begin{array}{cc} 1 + x_5 & x_1 + \xi
\\ x_1 - \xi & 1 - x_5
\end{array} \right) \label{510} . \eeq

The right action $ Sp(1) : S^7 \times Sp(1) \rightarrow S^7 $
transforms the vector $ | \psi > $ in a multiplicative manner

\beq | \psi > \rightarrow | \psi^w > = \left( \begin{array}{c} aw \\
bw \end{array} \right) = | \psi > w \ \ \ \ \ \ \forall w \in
Sp(1) \label{511}\eeq

while the projector $p$ remains invariant and therefore its
elements belong to the algebra $A_H$ instead of $B_H$, as it
should be.

The canonical connection associated with the projector $ \nabla =
p \cdot d $ has curvature given by

\beq \nabla^2 = p ( dp )^2 = | \psi >< \psi | d \psi><\psi|
d\psi><\psi| + | \psi >< d\psi | d\psi >< \psi | \label{512} .
\eeq

Because of the fact that $ < \psi | d \psi > $ is a $1$-form with
values in $H$, the first term is non-vanishing. The associated
Chern classes are

\begin{eqnarray}
& \ & C_1 (p) = - \frac{1}{2\pi i} Tr ( p (dp)^2 ) \nonumber \\
& \ & C_2 (p) = - \frac{1}{8 \pi^2} [ Tr ( p (dp)^4 ) - C_1(p)^2 ]
\label{513} .
\end{eqnarray}

Since the two-form $ p (dp)^2 $ has values in the purely imaginary
quaternions, its trace is vanishing:

\beq C_1(p) = 0 \label{514} . \eeq

Instead for the second Chern class we obtain

\beq C_2 (p) = - \frac{3}{8 \pi^2} d ( vol ( S^4 ))
\label{515}\eeq

and the corresponding Chern number is given by

\beq c_2 (p) = \int_{S^4} C_2 (p) = - \frac{3}{8 \pi^2} \int_{S^4}
d ( vol ( S^4 )) = - \frac{3}{8 \pi^2}  \frac{8 \pi^2}{3} = - 1
\label{516} . \eeq

To obtain a nonequivalent projector it is enough to take the
transpose of $p$

\beq q = \left( \begin{array}{cc} | a |^2 & b \overline{a}
\\ a \overline{b} & | b |^2
\end{array} \right) = \frac{1}{2} \left( \begin{array}{cc} 1 + x_5 & x_1 - \xi
\\ x_1 + \xi & 1 - x_5
\end{array} \right) \label{517}\eeq

but to write this projector in a ket-bra combination we need to
introduce a trick, see \cite{8} for details. In this case the Chern number
is equal to $1$. Having different topological charges the projectors $p$ and
$q$ are inequivalent.

It is easy to compute the $1$-form connection, associated with the projector $p$

\beq A^\nabla = < \psi | d \psi > = \overline{a} da + \overline{b}
db \label{518} . \eeq

Since $A^\nabla$ is anti-hermitian, it has values in the purely
imaginary quaternions that can be identified with the Lie algebra
$Sp(1) \simeq SU(2)$.

Non-equivalent gauge connections are obtained acting on the vector
$|\psi >$ with an element $ g \in GL(2,H) $ module $Sp(2) \simeq
Spin (5)$:

\beq | \psi > \rightarrow | \psi^g > = \frac{1}{{[ < \psi |
g^{\dagger} g | \psi > ]}^{1/2}} g | \psi > \label{519} . \eeq

Requiring that these transformations maintain the self-duality
condition of the instanton, $GL(2,H)$ must be reduced to the pure
conformal transformations. The ( preserving orientation )
conformal group of $S^4$ is $SL(2,H)$ and since

\beq dim ( SL(2,H)) - dim ( Sp(2) ) = 15 - 10 = 5 \label{520}\eeq

we obtain a five parameter family of instantons, i.e. exactly
the $ADHM$ construction of instantons.

At a non-commutative level this construction must be modified
since the coordinates of the fuzzy sphere do not form a closed
algebra, and we must add the contribution of the
$\hat{w}_{\mu\nu}$ coordinates. However we now show how to
construct projectors which tend with continuity to the $SU(2)$
projectors on $S^4$, in which the contribution of the extra
coordinates $\hat{w}_{\mu\nu}$ decouples in the classical limit.

To obtain that, we rewrite the quaternions $ ( a, b ) $ in terms
of the basic oscillators of the theory, see (\ref{43}):

\beq a =  \left( \begin{array}{cc} a_0 & - a_1^{\dagger} \\
a_1 & a^{\dagger}_0 \end{array} \right) \ \ \ \ \ \ \ \
b = \left( \begin{array}{cc} a_2 & - a_3^{\dagger} \\
a_3 & a^{\dagger}_2 \end{array} \right)  \label{521}\eeq

from which it follows that the combination $a\overline{b}$ is a
function of the fuzzy four-sphere coordinates, while the
combination $\overline{b} a$ is outside from the algebra $a_i
a^{\dagger}_j$. To succeed in obtaining combinations of the type (
$ a \overline{a}, a \overline{b}, b \overline{a}, b \overline{b} $
) in the projectors we must start from a vector of the form

\beq | \psi_0 > = \left( \begin{array}{c} a \\ b
\end{array} \right) \label{522} . \eeq

Imposing the normalization condition

\beq < \psi_0 | \psi_0 > = ( \overline{a} \overline{b} ) \left(
\begin{array}{c} a \\ b
\end{array} \right) = \overline{a} a + \overline{b} b = \left( \begin{array}{cc} \hat{N} & 0
\\ 0 & \hat{N} + 4 \end{array} \right) \label{523}
\eeq

we obtain a diagonal matrix with elements dependent on the number
operator $\hat{N}$, hence we simply redefine $|\psi_0 >$ as :

\beq | \psi_0 > \rightarrow | \psi > = \left( \begin{array}{c} a'
\\ b'
\end{array} \right) \ \ \ \ \ a' = a \sqrt{ h ( \hat{N} ) } \ \ \
b' = b \sqrt{ h ( \hat{N} ) } \label{524}\eeq

where

\beq h(\hat{N}) = \left( \begin{array}{cc} \frac{1}{\hat{N}} & 0
\\ 0 & \frac{1}{\hat{N} + 4} \end{array} \right) \label{525} . \eeq

We can develop the projectors as

\beq p_N = | \psi >< \psi | = \left( \begin{array}{c} a
\\ b
\end{array} \right) \left( \begin{array}{cc} h( \hat{N})
& 0
\\ 0 & h( \hat{N} ) \end{array} \right) ( \overline{a}
\overline{b} ) = \left( \begin{array}{cc} a h( \hat{N})
\overline{a}  & a h( \hat{N}) \overline{b}
\\ b h( \hat{N})
\overline{a} & b h( \hat{N} ) \overline{b}  \end{array} \right)
\label{526} . \eeq

Now these elements are functions not only of the coordinates
$\hat{x}_\mu$ but also of the $\hat{w}_{\mu\nu}$. For example
let's work out the first entry:

\begin{eqnarray}
& \ &  a h ( \hat{N} ) \overline{a} = \left( \begin{array}{cc} a_0 & - a_1^{\dagger} \\
a_1 & a^{\dagger}_0 \end{array} \right) \left( \begin{array}{cc} \frac{1}{\hat{N}} & 0 \\
0 & \frac{1}{\hat{N}+4} \end{array} \right)
\left( \begin{array}{cc} a^{\dagger}_0 &  a_1^{\dagger} \\
- a_1 & a_0 \end{array} \right) \nonumber \\
& \ &  =  \left(
\begin{array}{cc} \frac{1}{\hat{N}+1} a_0 a_0^{\dagger} +
\frac{1}{\hat{N}+3} a_1^{\dagger} a_1 & \ ( \frac{1}{\hat{N}+1} -
\frac{1}{\hat{N}+3} )
a_0 a_1^{\dagger} \\
 ( \frac{1}{\hat{N}+1} - \frac{1}{\hat{N}+3}) a_1 a_0^{\dagger}&
\frac{1}{\hat{N}+1} a_1 a_1^{\dagger} + \frac{1}{\hat{N}+3}
a_0^{\dagger} a_0 \end{array} \right) \label{527} . \end{eqnarray}

At this level the projector is no more singular in the number
operator $\hat{N}$ and it is possible to substitute to $\hat{N}$
its eigenvalue $n$. In this way we obtain the final form of the
projector $p_N$.

Its trace is determined by the formula

\begin{eqnarray}
Tr \ p_N & = & \left( \frac{n+4}{n+1} + \frac{n}{n+3} \right)\  Tr
I =
\left( \frac{(n+2)(n+3)(n+4)}{6} + \frac{n(n+1)(n+2)}{6} \right) \  = \nonumber \\
& \ & = \ 2 \ Tr I + ( n+2 ) \ < \ Tr I_P \ = \ 4 \ Tr I
\label{528}
\end{eqnarray}

and it is always an integer, for every value of $n$.

\section{Projectors and equations of motion}

Let's recall the most general Y-M action on the fuzzy four-sphere
in terms of matrix models:

\begin{eqnarray}
& \ &  S( \lambda ) = - \frac{1}{g^2} Tr \left[ \frac{1}{4} [
X_\mu, X_\nu ][ X_\mu, X_\nu ] + \frac{2 \lambda}{5(n+2) \rho}
\epsilon^{\mu\nu\lambda\rho\sigma} X_\mu X_\nu X_\lambda X_\rho
X_\sigma + \right. \nonumber \\
& \ & \left. + 8 ( 1 - \lambda ) \rho^2 X_\mu X_\mu \right]
\label{61} .
\end{eqnarray}

The matrix variable

\beq X_\mu = \rho ( \hat{G}_\mu + A_\mu ) \label{62}\eeq

is related to the fluctuation $A_\mu$, which contains the degrees of freedom of a pure Y-M
connection on a sphere $S^4$. The corresponding equations of motion

\begin{eqnarray}
& \ & [ X_\nu, [ X_\mu, X_\nu ]] + \frac{2 \lambda}{(n+2) \rho}
\epsilon^{\mu\nu\lambda\rho\sigma} X_\nu X_\lambda X_\rho X_\sigma
+ \nonumber \\
& \ & + 16 ( 1 - \lambda ) \rho^2 X_\mu = 0 \label{63}
\end{eqnarray}

are of course solved by the background $ X_\mu = \rho \hat{G}_\mu
$ due to the identity:

\begin{eqnarray}
& \ & [ G_\nu, [ G_\mu, G_\nu]] = - 16 G_\mu \nonumber \\
& \ & G_\mu = \frac{1}{8(n+2)} \epsilon^{\mu\nu\lambda\rho\sigma}
G_\nu G_\lambda G_\rho G_\sigma \label{64} .
\end{eqnarray}

Our aim is to prove that the projectors, introduced in section 4,
are solutions of a particular class of models $S(\lambda)$.

We firstly attempt to link the projector $p$ to the matrix
variable $X_\mu$ according to the formula

\beq \tilde{X}_\mu = \rho p  \hat{G}_\mu  p \label{65} . \eeq

In this naive identification we obtain a gauge invariant matrix variable, which
is not directly coincident with the standard one $X_\mu$. However by using the
property that the projector can be put in the form of a ket-bra function,

\beq p = | \psi >< \psi | \label{66}\eeq

the passage from the gauge invariant formulation $\tilde{X}_\mu$ to the usual
gauge covariant one $X_\mu$ is straightforward

\beq X_\mu = < \psi | \tilde{X}_\mu | \psi > =  < \psi | \rho
\hat{G}_\mu | \psi > = \rho \hat{G}_\mu + \rho < \psi | [
\hat{G}_\mu , | \psi > ] \label{67} . \eeq

At this level we recognize the fluctuation $A_\mu$ represented in terms of the
vector valued function $ | \psi > $:

\beq A_\mu = < \psi | [ \hat{G}_\mu , | \psi > ] \label{68} . \eeq

In ref. \cite{6} it was already noticed that introducing the link (\ref{65}) directly
in the classical action produces an ambiguity in the variational problem.
It is clear that the variation of the classical action with respect to a generic
projector $p$, subject only to the conditions $ p^2 = p,
p^{\dagger} = p $, is more general than the variation with respect to the
connection $\tilde{X}_\mu$.

To avoid such ambiguity we will discuss the link (\ref{65}) only for the $Y-M$ equations
of motion, showing that the instanton projectors (\ref{411}) and (\ref{415}) are indeed
solutions to them.

By introducing the link (\ref{65}), the equation of motion
(\ref{63}) can be written, thanks to the identity

\beq [ X_\mu, X_\nu ] = \rho^2 p ( [ G_\mu, p ] [ G_\nu, p] - [
G_\nu, p ] [ G_\mu, p] + [ G_\mu, G_\nu ] ) p \label{69}\eeq

as

\begin{eqnarray}
& \ & p [ G_\nu, ( [ G_\mu, p ][ G_\nu, p ] - ( \mu
\leftrightarrow \nu ) ) ] p + \nonumber \\
& \ & p [ [ G_\nu, p ], [ [ G_\mu, G_\nu ], p ] ] + \nonumber \\
& \ & \frac{\lambda}{2(n+2)} \epsilon^{\mu\nu\lambda\rho\sigma} (
4 p [ G_\nu, p ][
G_\lambda, p ] [ G_\rho, p ] [ G_\sigma, p ] + \nonumber \\
& \ & 2 p [ G_\nu, p ] [ G_\lambda, p ] [ G_\rho, G_\sigma ] p +
\nonumber \\
& \ & 2 p [ G_\nu , G_\lambda ] [ G_\rho, p ] [ G_\sigma, p ] p +
\nonumber \\
& \ & p [ [ G_\mu , G_\lambda ] , p ] [ [ G_\rho , G_\sigma ] , p
] ) = 0 \label{610} .
\end{eqnarray}

To show that the equation (\ref{610}) is solved by the projectors
$p_k$, it is simpler to compute directly $\tilde{X}_\mu$

\begin{eqnarray}
 \tilde{X}^{(k)}_\mu & = & \rho p_k  G_\mu p_k = \rho \frac{N}{ N+k } | \psi_k > G_\mu <
\psi_k | =   f_{(k)} ( \rho ) | \psi_k > G_\mu < \psi_k | \ \ \ \ k > 0 \nonumber \\
\tilde{X}^{(-k)}_\mu & = & \rho p_{(-k)}  G_\mu p_{(-k)} = \rho
\frac{ N+4 }{ N+4-k } | \psi_{(-k)} > G_\mu < \psi_{(-k)} | =
f_{(-k)} ( \rho ) | \psi_{(-k)} > G_\mu < \psi_{(-k)} | \nonumber
\\ & \ & 0 < k < N+1 \label{612} .
\end{eqnarray}

We can extract from (\ref{612}) the contribution of the ( gauge-invariant ) connection,
related to the charge $k$ of the instanton according to

\begin{eqnarray}
| \psi_k > A^{(k)}_\mu < \psi_k | = -\frac{ k }{ N+k } | \psi_k
> G_\mu < \psi_k | \ \ \ \ \ k > 0 \nonumber \\
| \psi_{(-k)} > A^{(-k)}_\mu < \psi_{(-k)} | = \frac{ k }{ N+4-k }
| \psi_{(-k)}
> G_\mu < \psi_{(-k)} | \ \ \ \ \ 0 < k < N+1
\label{613} . \end{eqnarray}

We notice that the solution produced by the projectors $p_k$ is a
simple re-scaling of the background, as it happens in the case of
non-commutative monopoles. By substituting the ansatz

\beq X_\mu^{\pm} = f_{\pm k} ( \rho ) \hat{G}_\mu \label{614}\eeq

in the equations of motion for the matrix variable $X_\mu$ we
obtain

\beq \lambda ( f_{\pm k} ( \rho )^2 + \rho f_{\pm k} ( \rho ) +
\rho^2 ) = \rho ( f_{\pm k} ( \rho ) + \rho ) \label{615}\eeq

from which we can fix the coupling constant $\lambda$ as a
function of $N$:

\begin{eqnarray}
& \ & \lambda = \frac{ \rho ( f_{\pm k} ( \rho ) + \rho ) }{
f_{\pm k} ( \rho
)^2 + \rho f_{\pm k} ( \rho ) + \rho^2 } \nonumber \\
& \ & f_{ k} ( \rho ) = \rho \frac{ N }{ N+k } = \rho ( 1 + O (
\frac{k}{N} )) \nonumber \\
& \ & f_{-k} ( \rho ) = \rho \frac{ N+4 }{ N+4-k } = \rho ( 1 + O
( \frac{k}{N} )) \label{616} .
\end{eqnarray}

The class of models, for which the noncommutative projectors $p_k$
are solutions, is of the type:

\beq \lambda = \frac{2}{3} + O ( \frac{k}{N} ) \label{617}\eeq

i.e. they are situated around the classical value $\lambda_{cl} =
\frac{2}{3}$. With a mechanism similar to the non-commutative
monopoles, we notice that to define non-commutative soliton
solutions on a fuzzy four-sphere it is necessary to perturb the
$\lambda$ coupling constant in a form

\beq \lambda = \lambda_{cl} + \frac{c}{N} \label{618} . \eeq

In the special case $\lambda = \frac{2}{3}$ we should obtain
soliton solutions related to the classical sphere $S^4$. However
the matrix model $X_\mu$ contains many extra degrees of freedom
with respect to the pure $Y-M$ theory, like the dependence on the extra
coordinates $w_{\mu\nu}$. We leave as an open question to check if these extra degrees of
freedom can be decoupled from the true variables of the classical
sphere $S^4$. We have already reached an important result i.e. we
know for what models our non-commutative projectors $p_k$ are
solutions of the $Y-M$ equations of motion.

\section{Reconstruction of the gauge connection from the
projectors}

The results obtained in section 6 are still partial, since to reconstruct the gauge
connection we cannot use directly the vector $ < \psi_k | $ as in (\ref{612}),
function of the oscillator algebra which is more general than the fuzzy
four-sphere algebra $a_i a^{\dagger}_j$. This fact produces a
discontinuity problem in the derivative action, since the adjoint
action of the background $[ \hat{G}_\mu, . ]$ on the oscillator
has the form:

\begin{eqnarray}
& \ & [ \hat{G}_1, . ] = \overline{a}_2 \frac{\partial}{\partial \overline{a}_0} -
a_0 \frac{\partial}{\partial a_2} + \overline{a}_1 \frac{\partial}{\partial \overline{a}_3} -
a_3 \frac{\partial}{\partial a_1} + \overline{a}_0 \frac{\partial}{\partial \overline{a}_2} -
a_2 \frac{\partial}{\partial a_0} + \overline{a}_3 \frac{\partial}{\partial \overline{a}_1} -
a_1 \frac{\partial}{\partial a_3} \nonumber \\
& \ & [ \hat{G}_2, . ] =  i ( \overline{a}_2 \frac{\partial}{\partial \overline{a}_0} -
a_0 \frac{\partial}{\partial a_2} + \overline{a}_1 \frac{\partial}{\partial \overline{a}_3} -
 a_3 \frac{\partial}{\partial a_1} - \overline{a}_0 \frac{\partial}{\partial \overline{a}_2} +
  a_2 \frac{\partial}{\partial a_0} - \overline{a}_3 \frac{\partial}{\partial \overline{a}_1} +
   a_1 \frac{\partial}{\partial a_2} ) \nonumber \\
& \ & [ \hat{G}_3, . ] = \overline{a}_3 \frac{\partial}{\partial \overline{a}_0} -
 a_0 \frac{\partial}{\partial a_3} - \overline{a}_1 \frac{\partial}{\partial \overline{a}_2} +
 a_2 \frac{\partial}{\partial a_1} + \overline{a}_0 \frac{\partial}{\partial \overline{a}_3} -
 a_3 \frac{\partial}{\partial a_0} - \overline{a}_2 \frac{\partial}{\partial \overline{a}_1}
 + a_1 \frac{\partial}{\partial a_2} \nonumber \\
& \ & [ \hat{G}_4, . ] = i ( \overline{a}_3 \frac{\partial}{\partial \overline{a}_0} -
a_0 \frac{\partial}{\partial a_3} - \overline{a}_1 \frac{\partial}{\partial \overline{a}_2} +
 a_2 \frac{\partial}{\partial a_1} - \overline{a}_0 \frac{\partial}{\partial \overline{a}_3} +
  a_3 \frac{\partial}{\partial a_0} + \overline{a}_2 \frac{\partial}{\partial \overline{a}_1} -
  a_1 \frac{\partial}{\partial a_2} )
\nonumber \\
& \ & [ \hat{G}_5, . ] = \overline{a}_0 \frac{\partial}{\partial \overline{a}_0} -
 a_0 \frac{\partial}{\partial a_0} + \overline{a}_1 \frac{\partial}{\partial \overline{a}_1} -
 a_1 \frac{\partial}{\partial a_1} - \overline{a}_2 \frac{\partial}{\partial \overline{a}_2} +
 a_2 \frac{\partial}{\partial a_2} - \overline{a}_3 \frac{\partial}{\partial \overline{a}_3} +
 a_3 \frac{\partial}{\partial a_3} \nonumber \\
 & \ &
  \label{71}
\end{eqnarray}

and it contains a dependence on a spurious angle which is
physically irrelevant because the physical operator algebra is
generated by the basic combinations $a_i a^{\dagger}_j$.

This problem has been solved in the case of non-commutative
monopoles \cite{26}, since the vector $ | \psi > $ has a gauge
arbitrariness $ | \psi > \rightarrow | \psi > U $, leaving the
projector invariant.

Consider for example the vector valued operator $ | \psi > $ in
the case $ k = - 1 $:

\begin{eqnarray}
& \ & | \psi_{k = -1} > = \frac{1}{ \sqrt{ \hat{N} + 3 }} \left(
\begin{array}{c} \overline{a}_0 \\ \overline{a}_1 \\ \overline{a}_2 \\
\overline{a}_3 \end{array} \right) = \left(
\begin{array}{c} \tilde{\overline{a}}_0 \\ \tilde{\overline{a}}_1 \\
\tilde{\overline{a}}_2 \\
\tilde{\overline{a}}_3 \end{array} \right) \label{72}
\end{eqnarray}

where

\begin{eqnarray}
& \ & \tilde{\overline{a}}_0 = \sum_{n_1 , n_2, n_3, n_4 =0}^{\infty}  \sqrt{
\frac{n_1 + 1 }{ n_1 + n_2 + n_3 + n_4 + 4 }} |
n_1+1, n_2, n_3, n_4 >< n_1, n_2, n_3, n_4 | \nonumber \\
& \ & \tilde{\overline{a}}_1 = \sum_{n_1 , n_2, n_3, n_4 =0}^{\infty}  \sqrt{
\frac{n_2 + 1 }{ n_1 + n_2 + n_3 + n_4 + 4 }} |
n_1, n_2+1, n_3, n_4 >< n_1, n_2, n_3, n_4 | \nonumber \\
& \ & \tilde{\overline{a}}_2 = \sum_{n_1 , n_2 , n_3 , n_4 =0}^{\infty} \sqrt{
\frac{n_3 + 1 }{ n_1 + n_2 + n_3 + n_4 + 4 }} |
n_1, n_2, n_3+1, n_4 >< n_1, n_2, n_3, n_4 | \nonumber \\
& \ & \tilde{\overline{a}}_3 = \sum_{n_1 , n_2 , n_3 , n_4 =0}^{\infty} \sqrt{
\frac{n_4 + 1 }{ n_1 + n_2 + n_3 + n_4 + 4 }} | n_1, n_2, n_3,
n_4+1 >< n_1, n_2, n_3, n_4 | \nonumber \\
& \ & < \psi_{-1} | \psi_{-1} > = 1 \label{73} .
\end{eqnarray}

Unfortunately we are facing with the problem that the action of $
| \psi_{-1} > $ doesn't commute with the number operator $ \hat{N}
$ and therefore it is not possible to restrict its action to a
fixed number $N$, as instead it is required for the construction
of a fuzzy four-sphere. We are going to redefine the vector $ |
\psi_{-1} > $, maintaining the projector $p_1$ invariant in form,
with an operator acting on the right and non-commuting with the
number operator, i.e. a quasi-unitary operator:

\beq | \psi_{-1} > \rightarrow | \psi_{-1}' > = | \psi_{-1} > U \
\ \ \ \ \ U U^{\dagger} = 1 \ \ \ \ ( U^{\dagger} U = 1 - P_0 )
\label{74} . \eeq

The condition $ U U^{\dagger} = 1 $ is enough to keep invariant
the form of the non-commutative projectors $p_k$ .

We will see that the presence of the quasi-unitary operator $U$
not only adjusts the classical limit but it reveals also the
topological character of the solution, at the matrix model level.
Many choices of $U$ are possible, for example:

\beq U_1 = \sum_{n_1 , n_2 , n_3 , n_4 =0}^{\infty}  | n_1, n_2, n_3, n_4 >< n_1+1,
n_2, n_3, n_4 | \label{75}\eeq

Interchanging $n_i \leftrightarrow n_j$ we obtain equivalent gauge
connections, related by a pure gauge transformation. Let us
compute

\begin{eqnarray}
& \ & | \psi_{-1}' > = | \psi_{-1} > U_1 = \left(
\begin{array}{c} \tilde{\overline{a'}}_0 \\ \tilde{\overline{a'}}_1 \\
\tilde{\overline{a'}}_2 \\
\tilde{\overline{a'}}_3 \end{array} \right) \nonumber \\
& \ & \tilde{\overline{a'}}_0 = \sum_{n_1 , n_2 , n_3 , n_4 =0}^{\infty} \sqrt{
\frac{n_1 + 1}{n_1+n_2+n_3+n_4 + 4}} |
n_1+1, n_2, n_3, n_4 >< n_1+1, n_2, n_3, n_4 | \nonumber \\
& \ & \tilde{\overline{a'}}_1 = \sum_{n_1 , n_2 , n_3 , n_4 =0}^{\infty} \sqrt{
\frac{n_2 + 1}{n_1+n_2+n_3+n_4 + 4}} |
n_1, n_2+1, n_3, n_4 >< n_1+1, n_2, n_3, n_4 | \nonumber \\
& \ & \tilde{\overline{a'}}_2 = \sum_{n_1 , n_2 , n_3 , n_4 =0}^{\infty} \sqrt{
\frac{n_3 + 1}{n_1+n_2+n_3+n_4 + 4}}|
n_1, n_2, n_3+1, n_4 >< n_1+1, n_2, n_3, n_4 | \nonumber \\
& \ & \tilde{\overline{a'}}_3 = \sum_{n_1 , n_2 , n_3 , n_4 =0}^{\infty} \sqrt{
\frac{n_4 + 1}{n_1+n_2+n_3+n_4 + 4}} | n_1, n_2, n_3, n_4+1 ><
n_1+1, n_2, n_3, n_4 |\nonumber \\
& \ & \label{76} .
\end{eqnarray}

Since, by construction, $ [ \hat{N}, | \psi'_{-1} > ] = 0 $ the
action of $ | \psi'_{-1} > $ is well defined at a fixed $N=n$,
i.e. for a particular fuzzy four-sphere:

\begin{eqnarray}
& \ & \tilde{\overline{a'}}_0|_N = \sum_{k_1 =0}^{N-1} \sum_{k_2
=0}^{N-k_1-1} \sum_{k_3=0}^{N-k_1-k_2-1}  \sqrt{\frac{
k_1+1}{N+3}} \nonumber \\
& \ & |
k_1+1, k_2, k_3, N-k_1-k_2-k_3-1 >< k_1+1, k_2, k_3, N-k_1-k_2-k_3-1 | \nonumber \\
& \ & \tilde{\overline{a'}}_1|_N = \sum_{k_1 =0}^{N-1} \sum_{k_2
=0}^{N-k_1-1} \sum_{k_3 =0}^{N-k_1-k_2-1}  \sqrt{\frac{
k_2+1}{N+3}} \nonumber \\ & \ & |
k_1, k_2+1, k_3, N-k_1-k_2-k_3-1 >< k_1+1, k_2, k_3, N-k_1-k_2-k_3-1 | \nonumber \\
& \ & \tilde{\overline{a'}}_2|_N = \sum_{k_1 =0}^{N-1} \sum_{k_2
=0}^{N-k_1-1} \sum_{k_3 =0}^{N-k_1-k_2-1} \sqrt{\frac{
k_3+1}{N+3}} \nonumber \\ & \ & |
k_1, k_2, k_3+1, N-k_1-k_2-k_3-1 >< k_1+1, k_2, k_3, N-k_1-k_2-k_3-1 | \nonumber \\
& \ & \tilde{\overline{a'}}_3|_N = \sum_{k_1 =0}^{N-1} \sum_{k_2
=0}^{N-k_1-1} \sum_{k_3 =0}^{N-k_1-k_2-1}  \sqrt{\frac{
N-k_1-k_2-k_3}{N+3}} \nonumber \\ & \ & | k_1, k_2, k_3,
N-k_1-k_2-k_3
>< k_1+1, k_2, k_3, N-k_1-k_2-k_3-1 | \label{77} .
\end{eqnarray}

These actions now belong to the functional space of the fuzzy
four-sphere and they can be recast in terms of the generalized
spherical harmonics.

Having solved the discontinuity problem in the derivative action,
it is straightforward to prove that $ | \psi'_{-1} > $ gives rise
to connections satisfying the $Y-M$ equations of motion ( in the
gauge-covariant formulation ) since

\beq \tilde{X}_\mu = | \psi_{-1} > U f_{-1}( \rho ) ( U^{\dagger}
G_\mu U ) U^{\dagger} < \psi_{-1} | = | \psi_{-1}' > X_\mu  <
\psi_{-1}' | \label{78} . \eeq

Now we can look at the physical solution $X_\mu$, of the form:

\beq X_\mu = f_{-1}( \rho ) U^{\dagger} G_\mu U \label{79}\eeq

that satisfies the matrix model equations of motion, if $\lambda$
satisfies to (\ref{616}). The generalization of these results to
the case $ | \psi_{-k} > $ with $k$ generic is direct.

Summarizing all these results, the solution of the matrix model
$X_\mu$ for charge $-k$ is therefore obtained in two steps:

i) firstly by re-scaling  the background solution $ X^{(0)}_\mu =
f( \rho ) G_\mu $ ;

ii) secondly by dressing $ X^{(0)}_\mu $  as $ X_\mu = U^{\dagger}
f( \rho ) G_\mu U $; the quasi-unitary operator $U$ maps the
background to a reducible representation of the fuzzy four-sphere
algebra, making clear the topological nature of the solution.

It remains to be investigated if this construction can be repeated
for the case $ | \psi_k > \ ( k > 0 )$, for example $k=1$

\begin{eqnarray}
& \ & | \psi_{k = 1} > = \frac{1}{ \sqrt{ \hat{N} + 1}} \left(
\begin{array}{c} a_0 \\ a_1 \\ a_2 \\
a_3 \end{array} \right) = \left(
\begin{array}{c} \tilde{a}_0 \\ \tilde{a}_1 \\
\tilde{a}_2 \\
\tilde{a}_3 \end{array} \right) \nonumber \\
& \ & \tilde{a}_0 = \sum_{n_1 , n_2 , n_3 , n_4 =0}^{\infty}
\sqrt{\frac{n_1+1}{n_1+n_2+n_3+n_4+1}} |
n_1, n_2, n_3, n_4 >< n_1+1, n_2, n_3, n_4 | \nonumber \\
& \ & \tilde{a}_1 = \sum_{n_1 , n_2 , n_3 , n_4 =0}^{\infty}
\sqrt{\frac{n_2+1}{n_1+n_2+n_3+n_4+1}} |
n_1, n_2, n_3, n_4 >< n_1, n_2+1, n_3, n_4 | \nonumber \\
& \ & \tilde{a}_2 = \sum_{n_1 , n_2 , n_3 , n_4 =0}^{\infty}
\sqrt{\frac{n_3+1}{n_1+n_2+n_3+n_4+1}} |
n_1, n_2, n_3, n_4 >< n_1, n_2, n_3+1, n_4 | \nonumber \\
& \ & \tilde{a}_3 = \sum_{n_1 , n_2, n_3 , n_4 =0}^{\infty}
\sqrt{\frac{n_4+1}{n_1+n_2+n_3+n_4+1}} | n_1, n_2,
n_3, n_4 >< n_1, n_2, n_3, n_4+1 | \nonumber \\
& \ & < \psi_{k=1} | = ( \overline{a}_0 , \overline{a}_1,
\overline{a}_2 , \overline{a}_3 ) \frac{1}{ \sqrt{
 \hat{N} + 1 }} = ( \tilde{\overline{a}}_0, \tilde{\overline{a}}_1,
 \tilde{\overline{a}}_2,
\tilde{\overline{a}}_3 ) \nonumber \\
& \ & < \psi_1 | \psi_1 > = 1 - < 0,0,0,0 | 0,0,0,0 > = 1 - P_0
\label{710} . \end{eqnarray}

In the last normalization condition we can forget the presence of
$P_0$, when verifying that $p_k$ is a projector, since

\beq | \psi_1 > P_0 = P_0 < \psi_1 | = 0 \label{711}\eeq

the action of $ | \psi_1  > $ on the projector $P_0$ is null.

However to redefine $ | \psi_1  > $  in order that $ [ N, \psi_1'
] $ is satisfied, it is necessary to use the adjoint of the
quasi-unitary operator $U_1$:

\beq U^{\dagger}_1 = \sum_{n_1 , n_2 , n_3 , n_4 =0}^{\infty}
   | n_1+1, n_2, n_3, n_4 >< n_1, n_2, n_3, n_4 | \label{712} . \eeq

Unfortunately the dressing with $U^{\dagger}_1$ changes the form
of the projector $p_k$ since

\beq | \psi'_1 >< \psi'_1 | = | \psi_1 > U^{\dagger} U < \psi_1 | =
   | \psi_1 > \left( 1 - \sum_{n_2 , n_3, n_4 =0}^{\infty}
    | 0 , n_2, n_3, n_4 >< 0 , n_2, n_3, n_4 | \right) < \psi_1 |
\label{713} . \eeq

There is an extra contribution in parenthesis not cancelled by the
presence of $ | \psi_1 > $. We conclude that it is impossible for
the charge $k$-instantons to define a connection satisfying the
$Y-M$ equations of motion while the corresponding projectors do
it, similarly to what we have found for the noncommutative
monopoles in Ref. \cite{26}.

\section{Conclusions}

In Ref. \cite{4} a finite module description of $U(1)$ instantons
on a fuzzy four-sphere has been presented. In this work we have
investigated the relationship between these projectors and the
matrix model, which defines the physical dynamics from a
connection point of view. Basically we have reached two results,
i.e. finding the physical models for which the projectors of Ref.
\cite{4} are solution to the corresponding equations of motion and
characterizing the nontrivial topology at the level of the matrix
model.

However it remains open the more interesting case of $U(2)$ gauge theory, leading in the
classical limit to the BPST $SU(2)$ instanton, which is complicated by the structure of
the quaternions. Already in the $U(1)$ case we have identified the main ingredients to
build a noncommutative topology on a fuzzy four-sphere:

i) the presence of a scaling factor, which allows for the smooth limit to a commutative
solution;

ii) the presence of quasi-unitary operators acting on the background of the matrix model,
which map it, an irreducible representation of the $SO(5,1)$ algebra, to reducible
representations.

In this sense we are able to find connections satisfying the $Y-M$ equations of motion only
for negative charge projectors $p_{-k} ( 0 < k < N+1 ) $, since in a matrix variable $X_i$
with fixed dimension we can insert only a ( reducible ) representation with rank less than
that of $X_i$ but not bigger than $X_i$.

We believe that the finite module description is a very powerful method in noncommutative
geometry, but we hope with this work to make a bridge with the more familiar physical
language of connections, which is encoded in the matrix model approach.

\end{document}